# Sea Ice Concentration Estimation Techniques Using Machine Learning: An End-To-End Workflow for Estimating Concentration Maps from SAR Images[1]

*Stefan Dominicus[2] and Amit Kumar Mishra[3]*

Sea ice concentration is an important metric used to characterize polar sea ice behavior. Understanding this behavior and accurately representing it is of critical importance for climate science research, and also has important uses in the context of maritime navigation. An end-to-end workflow for generating learned concentration estimation models from synthetic aperture radar data, trained on existing passive microwave data, is presented here. A novel objective function was introduced to account for uncertainty in the passive microwave measurements, which can be extended to account for arbitrary sources of error in the training data, and a recent set of in-situ observations was used to evaluate the reliability of the chosen passive microwave concentration estimation model. Google Colaboratory was used as the development platform, and all notebooks, training data, and trained models are available on GitHub. This chapter is an overview of the most interesting aspects of this investigation, and a detailed report is also available on GitHub.

*GitHub Repository Link:*
*https://github.com/stefandominicus/FYP_ML_SIC*

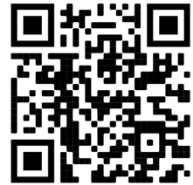

## 9.1.    Introduction

Polar sea ice coverage, and the phenomena which drive changes in this coverage, form a critical link in the Earth's climate system. Sea ice is located right at the ocean-atmosphere interface and is thus intricately linked to climate system evolution [1]. Current models used to estimate sea ice concentration (SIC) are derived from passive microwave (PMW) radar satellite data, but these instruments offer poor spatial resolutions, and are susceptible to interference due to atmospheric variations. Additionally, these models are usually calibrated for Arctic conditions, and evidence from recent in-situ observations in the Southern Ocean suggests that their estimates are significantly biased and are not reliable in the Antarctic marginal ice zone (MIZ). Sea ice behavior is characterized by a number of important variables, of which SIC is one, which have each been tied to particular aspects of climate system evolution. It is clear that understanding the interplay between these variables is an important requirement for accurately modelling large-scale climate trends [2].

---

[2] Honors Student, Electrical Engineering Department, University of Cape Town, South Africa, stefandominicus@gmail.com
[3] Professor, Radar Remote Sensing Group, University of Cape Town, South Africa akmishra@ieee.org

### 9.1.1.  Context

Ice edge location has been the traditional choice for quantifying ice coverage and is defined as the point where the ice concentration first exceeds 15% [2]. However, recent investigations in the Southern Ocean have chosen to focus on the much broader Marginal Ice Zone (MIZ), which is defined as the region for which the sea ice concentration lies between 15% and 80%. Studying concentration rather than edge location has highlighted many of the shortcomings in current modeling techniques [3].

There are a few factors which make this a difficult problem to solve. First, consider that much of what is known in the Arctic in terms of climate dynamics, and sea ice in particular, has been shown to be of little use in the Antarctic region due to significant differences in the underlying driving forces influencing sea ice dynamics in each region. There is currently no consensus on what factors may be the primary driving forces of sea ice dynamics in the Antarctic [2]. Additionally, much of the current understanding in this field regarding sea ice characteristics was derived from old Arctic sea ice, which is of questionable relevance to the present day Arctic environment, let alone the Antarctic [4].

The second, and perhaps more critical factor hindering the development of accurate Antarctic sea ice models is the chronic lack of validated data. Although more in situ measurement data is slowly becoming available, the need still exceeds what existing datasets have to offer. The accuracy of existing models may be sufficient for some long-term synoptic analysis, but coupled models are still not able to capture large scale sea ice coverage features due to incomplete parameterizations of the system dynamics [4].

Satellite-based remote sensing has been proposed as a way to obtain more data from the polar regions, but perhaps the biggest obstacle when using satellite data is that some understanding of the sensed environment is still required in order to properly interpret the data provided by the remote platform [5]. Interpretation models have been developed for the Arctic region, aided by the superior availability of validation data. As before, these models can be transferred for use in the Antarctic, but will not capture many important details [3], which renders the model all but useless when a high temporal resolution is desired [1].

Another factor which has to some extent precluded the use of satellite observations in Antarctic sea ice analysis is that so much of the characteristic behavior takes place during intense stormy conditions, where clouds would obscure the scene from view by a satellite [3]. Even if passive microwave (PMW) instruments are used, the increased water and water vapor content in the atmosphere can significantly degrade the quality and accuracy of the reported information [6]. However, a promising alternative is to use Synthetic Aperture Radar (SAR) instruments which are not affected by cloud cover or sunlight and are much less sensitive to atmospheric water vapor content. The use of SAR platforms also presents the advantage of increased spatial resolution [4]. This improvement in spatial resolution will allow the interpretive models being developed to achieve greater intricacy [1].

## 9.1.2. *Existing work*

In this context, there is a clear desire for an improved modelling approach which is not reliant on existing numerical models, can offer improved spatio-temporal resolutions, is flexible and can be adapted to suit the region of interest, and which takes into account the uncertainty present in any datasets used so as not to be hindered by the high variability of some existing measurements. A recent paper presented a novel method of sea ice concentration (SIC) estimation using deep learning [6], and was the most comprehensive effort found in addressing these desires.

Traditionally, when higher spatial resolution information is required, it can be obtained by the manual interpretation of SAR images, which also have the advantage of being less susceptible to interference from atmospheric water content (compared to the more widely used PMW data). While this method is certainly able to produce accurate interpretations, it is very time consuming, and simply cannot keep up with the vast number of SAR images generated.

While deep learning is indeed well-suited to this kind of high-volume data processing task, previous efforts have relied on manually annotated SAR images for training, which are generally only available in specific regions, and only describe SIC in discrete 10-20% concentration bands, which leads to coarse and biased interpretations. In [6], SIC information generated automatically from PMW data was used to generate concentration labels corresponding to a set of SAR images. These image-label pairs were then used to train a deep neural network, based on the DenseNet architecture [7].

It is worth discussing some of the challenges, design choices, and outcomes presented in [6], as they were a major influence on the path taken in this investigation. Firstly, the size of SAR images (around 10,000 x10,000 pixels) is prohibitively large to be used directly in neural networks and was addressed by breaking each image into smaller patches which were each processed individually. Secondly, the difference in resolution between the SAR images and the existing SIC data necessitated the use of an interpolation and resampling scheme to generate a concentration label corresponding to the footprint of each SAR image. Thirdly, not all of the available image patches were used for training - only those which contained a representative range of SIC values were included in the training dataset. In terms of neural network architecture, the popular DenseNet structure was used with simple modifications made to the output to allow for single value regression, rather than classification for which the architecture is already well-known. Lastly, this approach presented evidence that models trained in this way were able to predict SIC features at a higher resolution than the data with which they were trained, and also suggested that some models would inherit the flaws and inaccuracies of the PMW data used for training, while others were able to avoid these flaws, and actually produce more accurate predictions than the data with which they were trained [6].

Much of the novelty in the original DenseNet structure was in the way it utilized skip connections to aid in gradient propagation during training, making larger networks more efficient to train [7]. One of the most well-known architectures to also leverage skip connections is U-Net [8], which was introduced one year prior to DenseNet. Where DenseNet was designed for classification, U-Net was designed for image segmentation tasks, and its encoder-decoder type structure is naturally suited to producing outputs of a similar dimension to the input. This is particularly interesting in this context, where the improved resolution of SAR images is perhaps one of the most important factors in trying to generate higher spatial resolution SIC maps. Additionally, the original U-Net implementation also used sample weighting maps in the training process, which served to emphasize particular features, and cause the network to learn those features above others [8]. This method of training manipulation by error-weighting presents some interesting opportunities in terms of allowing the training process to account for known inaccuracies in the training data and may play an important role in allowing learned models to avoid inheriting the flaws present in the models used to generate training data.

### 9.1.3.   Major contributions

The work presented here is part of a more extensive project, with the intention of presenting the most interesting outcomes in a more concise manner. The full report and more details about the project can be found on GitHub. For the purposes of this chapter, here are the major contributions which will be discussed.

- A fully cloud-based workflow for machine learning research and prototyping
- Building interactive tools in Google Collaboratory for ease of use
- Identification of an error in the Copernicus image data processor
- Insight into the benefits of multi-stage/mixed-data training schemes
- Evaluation of existing concentration estimates using in-situ observations

### 9.1.4.   Chapter structure

The development process in this investigation can be broken into three broad sections:
1. data pre-processing,
2. neural network development, and
3. serving targeted predictions.

The full details of each of these sections can be found in the project GitHub repository, but for the purpose of this chapter, the discussion will cover an overview of the data processing pipeline, a more explanation of the filtering, batching, and validation processes used, followed by an overview of the machine learning models and custom objective function used for this investigation. Some of the major contributions are then discussed in more detail, justifying their significance in this context. Lastly, some brief conclusions are presented, as well as recommendations for continued work in this direction.

## 9.2.    Data Pipeline Design

Every stage of the processing pipeline, machine learning development, and model evaluation described in this chapter was built online using Google Colaboratory running Python 3. This presents a number of useful advantages, such as free access to GPU hardware accelerators for machine learning development. But is not without its challenges, such as limited runtime and issues around storage virtualization. Even so, the speed increase offered by the use of GPUs compared to CPUs is so significant that Colaboratory was an obvious choice in the absence of any other available hardware accelerators. Much of the focus during development was to produce an end-to-end pipeline capable of demonstrating that free cloud platforms can in fact be used for more comprehensive tasks. The solutions and workarounds presented here can be easily adapted to other projects and will hopefully make free cloud computing more accessible for prototyping and research. The specific Python modules used at each stage will be mentioned as they are used, but for now it is worth mentioning that all image processing was done using *OpenCV*, *Matplotlib* was used to display all results, and *Keras* was used for all machine learning development, using the *TensorFlow* backend.

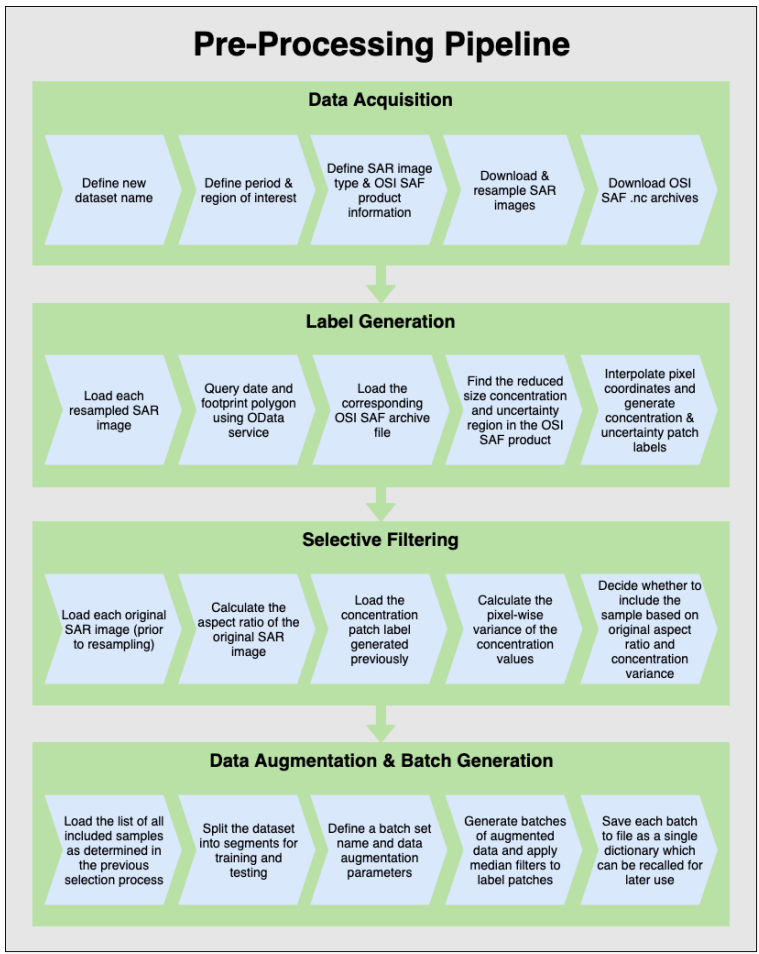

*Figure 1: Summary of the data pre-processing pipeline developed for this investigation. The pipeline was divided into four major stages, and this graphic illustrates how each stage links together, as well as how data moves through each stage in the pipeline.*

One of the complications which arise when trying to work with machine learning using an online development environment such as Colaboratory is the need to access large datasets for training each model. While premium services like Google Cloud or AWS EC2 are highly customizable, free services like Colaboratory always spin up as a fresh instance, and any previous work will have been lost. To make the most of this free service, a Google Drive folder was mounted as a remote storage device to each new instance. All files were then saved into this remote drive so that work could continue from one instance to the next. This method of data management has some significant speed limitations when dealing with larger datasets, which had to be accommodated in the processing pipeline. More discussion on this issue will follow below.

## 9.3.    Data Processing & Validation

### 9.3.1.   Creating a Labeled Dataset

In supervised machine learning, much of the challenge often lies in finding or creating suitable datasets on which to train the machine learning models. In this case, the objective was to generate concentration maps from SAR images. Training data was generated from an existing PMW-derived SIC model. The resulting dataset contained image-label pairs, where images are resampled dual polarization SAR measurements, and labels are the concentration and uncertainty patches corresponding to the SAR image footprint for the same day. In this investigation, SENTINEL-1 SAR data was chosen, as it has good coverage over polar regions and is freely available for research use. The PMW-derived SIC product chosen is published by OSI SAF, and designated OSI-401-b. This product has global coverage, meaning that its concentration maps cover both polar regions fully. Since each SAR individual image has a much smaller coverage, a resampling and interpolation algorithm was devised which could generate concentration and uncertainty patches from the global SIC chart which matched the location and resolution of the SAR image. This process of automatically generating label data was crucial for this methodology, as training was not limited to a handful of manually interpreted samples as has often been the case in previous work.

### 9.3.2.   Selective Filtering

Previous work using SAR images for machine learning have usually selected a handful of images with which to work, and in doing so it is ensured that the data being learned is in some way representative of the property of interest. However, in this experimental setup, *all* SAR images from the region of interest were downloaded. This meant that many of the images contained entirely 0% concentration (open water), or 100% concentration (consolidated ice or land). These images are of no interest, since they do not contain any variation in ice concentration and will skew the training process in favor of single-value concentration patch estimations, which is undesirable.

The solution proposed in this investigation is to filter the dataset prior to training and retain only images which contain interesting variations in ice concentration such as at the ice edge where the concentration values in a single image span most of the 0% - 100% range. A histogram was plotted for a subset of the labelled images, which is shown below in Figure 2. Pixel-wise variance in the concentration patch label was used to quantify the spread of concentration values within each image, and a threshold was used to filter the dataset based on the concentration variance – images with a suitably high variance were marked as acceptable, while the others were excluded from all further use.

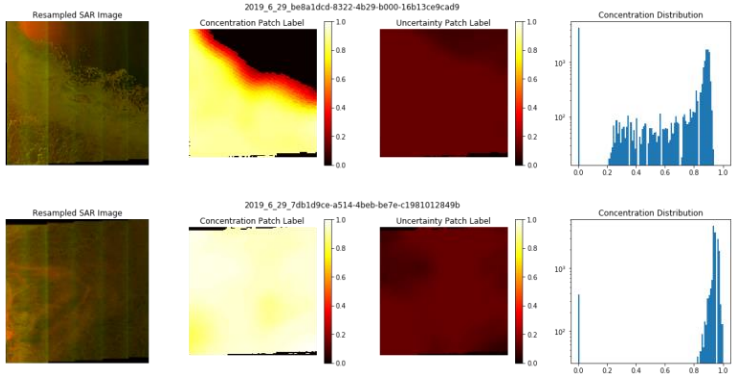

*Figure 2: Pixel-wise concentration distribution. In this figure, the resampled SAR image, concentration patch label and uncertainty patch label are shown for two representative samples in the dataset. The histogram on the right shows the distribution of concentration values present in each concentration patch label. The first sample (top) is clearly more "interesting" in terms of concentration features, since it includes a section of ice edge, while the second sample (bottom) contains almost no variation, with most of the concentration values being greater than 80%. A variance threshold was used to filter all samples, ensuring only "interesting" were included.*

## 9.3.3. Batching & Augmentation

*Keras* provides a number of helpful tools for batch training on image data, and the *ImageDataGenerator* class is particularly powerful when it comes to feeding batches into a neural network. All tests were conducted both with and without data augmentation (random flips and rotations) to understand the efficacy of the augmentation process, and a 5x5 median filter was also applied to all concentration and uncertainty patches. This filter helped to smooth out the rather grainy patch labels, while preserving edges, as shown in Figure 3 below. The granularity in the unfiltered patches is due in part to the poorer 10km resolution offered by the OSI SAF concentration chart used, and the median filter applied had the desired effect of smoothing the patch without blurring.

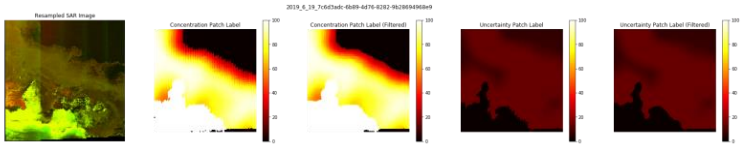

*Figure 3: Median filter (5x5) applied to concentration and uncertainty labels. The median filter does well at smoothing the patch labels without blurring the image, and also doesn't introduce any concentration values which were not already present in the vicinity of each pixel to begin with.*

Google Collaboratory's storage virtualization imposes a significant bottleneck when trying to access so many files individually. To reduce the number of fill access calls and directory jumps during training, the newly labelled dataset was processed into batches, and each batch was saved as a single file. With a batch size of 16, this reduced the number of file access calls by a factor of 48, which successfully avoided starving the GPU while training. Recalling batches of input and label data like this was facilitated by a extending the *Keras Sequence* class.

### 9.3.4.   Evaluation Interfaces

While existing sea ice concentration chart products, such as OSI-401-b, tend to focus on achieving global coverage in a single chart, the aim of this investigation was to produce a more focused concentration estimation for specific regions on demand. This kind of product is much more useful when providing close navigational support for shipping and research activities. A boilerplate search interface was developed in Google Colaboratory using *Forms* allowing a user to search for SAR images covering a particular location and date range. A screenshot of the Form interface and an arbitrary result set is shown in Figure 4 below.

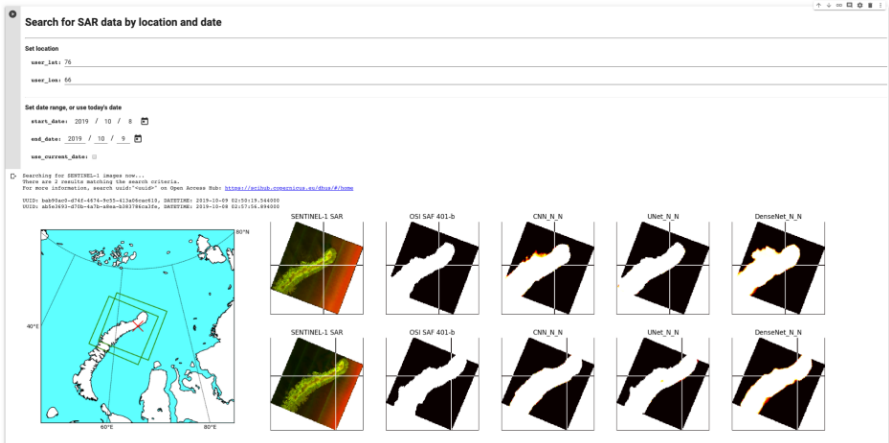

*Figure 4: This search interface allows a user to find and compare SAR images with concentration estimates from a variety of models side-by-side, with additional context from the map on the left-hand side.*

The map plot shown in Figure 4 above was dynamically generated at runtime and is centered on the specified location of interest. This interface was a very useful tool when developing the data pre-processing pipeline, as it allowed for specific edge cases to be visually inspected and validated. It also provided a very direct way to visually compare the concentration estimations generated by each model. Lastly, this interface offers a proof of concept for the intended workflow in the context of serving targeted predictions for close navigational support.

A secondary objective for this investigation was to compare estimates made by the existing numerical sea ice concentration models with a set of recent in situ observations. A 2019 research expedition to the MIZ in the Southern Ocean collected a large set of geotagged images of the sea ice conditions [9]. This kind of comparison is ideal for understanding the validity of the existing models in the Southern Ocean where extensive in situ observation data is lacking. Figure 5 shows a screenshot of the interface used to compare OSI-401-b predictions with in-situ observations of sea ice conditions.

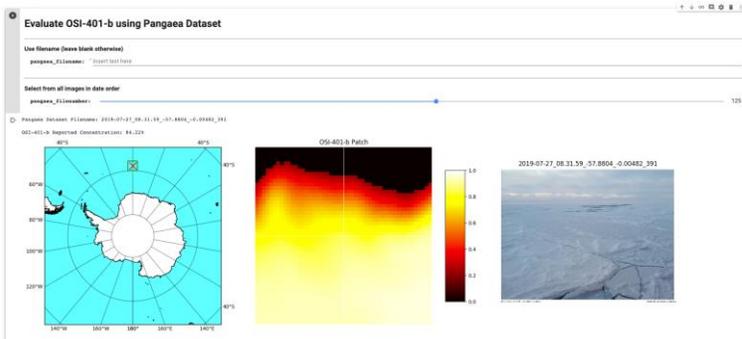

*Figure 5: This comparison interface shows images taken in the MIZ and compares them with the PMW-derived estimates for the same day and location, with the exact location indicated with crosshairs.*

### 9.3.5. Pass Direction Error

The search interface shown in the previous section rotated each SAR image so that it appears aligned with the global reference frame. However, after implementing this rotation and footprint overlay process, some images still appeared to be incorrectly rotated. Extensive troubleshooting revealed that the cause was an inconsistent mislabeling of some SAR products' pass direction attribute, which caused the image to be pre-processed incorrectly by Copernicus.

While troubleshooting this issue, it was noticed that the product *OData* '*footprint*' attribute (a set of points defining the footprint polygon) follows a predictable structure: the order in which the points are listed is always top-left, top-right, bottom-right, bottom-left (in the satellite's frame of reference). This is depicted in Figure 6 below.

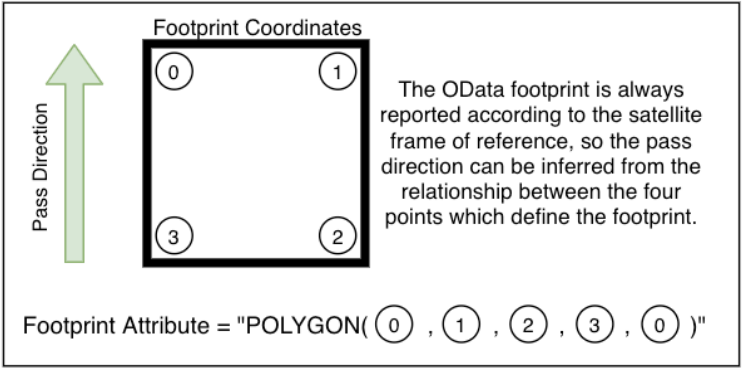

*Figure 6: A diagram explaining the OData 'footprint' structure. This can be leveraged to determine the satellite's actual orientation at acquisition time, without relying on the 'pass direction' attribute.*

This insight allows the actual pass direction to be determined without relying on the attribute reported by either *OData* or *OpenSearch*, which are both unreliable. One way to do this is to compare the latitudes of points 0 and 3. If the latitude of point 0 is greater than that of point 3, then the satellite was North facing at acquisition time, and the product is therefore ASCENDING. If not, then the product is DESCENDING. This derived pass direction was then compared to the reported *OData* attribute. If the two agree, then the pass direction attribute is correct, and the Quicklook image has been correctly pre-processed before being downloaded. If not, then the product was incorrectly labelled, and therefore the Quicklook image was incorrectly pre-processed before being downloaded. This erroneous processing could then be easily corrected, and further processing can continue as usual.

This information was reported to the Copernicus team for review and has been acknowledged as a known issue. It has been labelled as a low priority issue since most research activity uses the full SAR measurement, not the Quicklook image, which is not processed in the same way. This enquiry has been noted as the first time this issue has been reported externally, and the Copernicus team has stated that they intend to implement a fix in the next processor revision.

## 9.4.    Architectures & Training Schemes

### 9.4.1.   Summary of Models Used

Three different neural network architectures were used in this investigation, with the intention of comparing the performance and complexity of each. Hyperparameters were iteratively tuned until the mean absolute error on unseen test data was below 10%. All neural network models were developed using *Keras*.

Before investigating any exotic model architectures, a relatively straightforward sequential Fully Convolutional Neural Network (FCNN) was implemented to act as a baseline with which to compare the predictions of more complex models implemented later. Each layer was comprised of a convolutional layer with $n \times (3 \times 3)$ filter kernels and ReLU activation functions, followed by a dropout layer with an adjustable dropout percentage. A simple two-layer version of this architecture is shown in Figure 7 below. The convolution operators were padded so as to maintain the same image resolution throughout the network, which meant that the output layer was simply a single convolutional filter with a Sigmoid activation so that all pixel values fall in the range (0,1) representing the estimated concentration percentage.

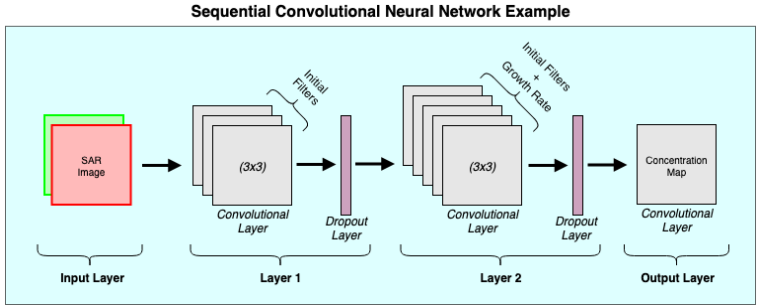

*Figure 7: An illustration of the sequential FCNN architecture. This example contains two hidden layers and illustrates how the number of convolutional filters is increased at each layer.*

The U-Net architecture was introduced in 2015 and has been successful in many image-segmentation tasks [8]. Since segmentation in this context is little more than pixel-wise classification, this architecture can be easily modified for concentration mapping, which is essentially pixel-wise regression. By modifying the final output layer, all the advantages of this network structure are retained, and very little design work is required. Figure 8 below shows the U-Net architecture as described in the original paper. A *Keras* implementation of the U-Net architecture found on GitHub was used for the bulk of the model definition [10].

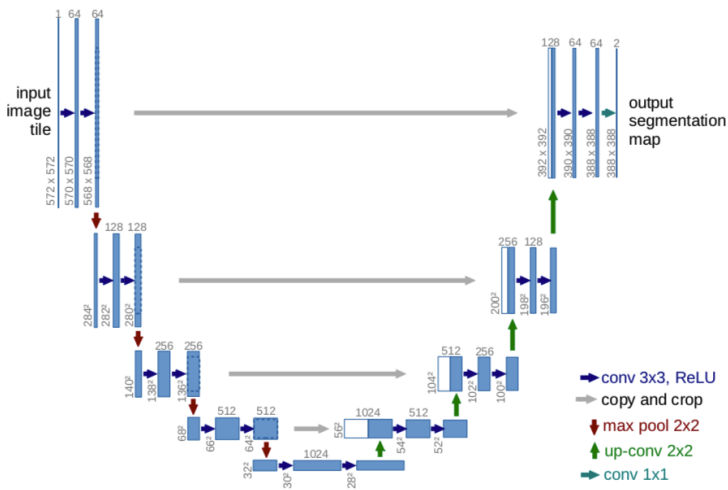

*Figure 8: A diagram illustrating the original U-Net architecture. Max pooling layers were removed so as to preserve the image dimension throughout the network. [8].*

Building on the success of skip connections as demonstrated in U-Net, another unique architecture was introduced where convolutional layers were densely connected to one another, called DenseNet [7]. The network was originally intended for use on classification problems, and Figure 9 below describes how the dense blocks (containing densely connected convolutional layers) are arranged sequentially with some transition operators in between each block. A Keras implementation of this architecture was used as a template for this investigation, found on GitHub [11]. Besides the changes to convolution padding and the removal of max pooling, the models weight decay hyperparameter was also set to zero. The purpose of this hyperparameter is to penalize excessively large weights in any of the model layers, in order to prevent overfitting to particular training examples, and overreliance on particular feature sets. However, this interfered with the training process by diluting the training loss value. So instead of minimizing the objective function by estimating more accurate concentration maps, the model training process simply minimized the layer weights, leading to very poor visual performance despite the training loss following the expected exponential decay trajectory.

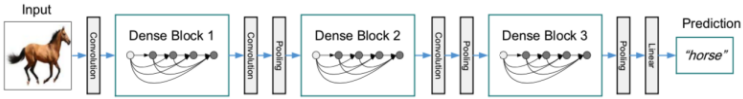

*Figure 9: A diagram illustrating the original DenseNet architecture. Although originally designed for classification, the structure was easily modified to predict the concentration maps needed in this investigation [7].*

For each of the model architectures described above, a number of variants were trained and evaluated. Each structure was iteratively tuned over a number of generations, until the estimation mean absolute error on unseen data was acceptable. The final configuration of each model is listed in Table 1 below.

| BASE ARCHITECTURE | STRUCTURE | DESCRIPTION |
|---|---|---|
| CNN | Layers: 10<br>Initial Filters: 32<br>Growth Rate: 32<br>Dropout Rate: 0.2<br>Parameters: 3,043,937 | A straightforward sequential CNN, where each layer is comprised of a convolutional layer followed by a dropout layer. |
| U-Net | Layers: 4<br>Initial Filters: 128<br>Growth Rate: x2<br>Dropout Rate: 0.5<br>Parameters: 124,108,545 | Based on the successful image segmentation network, this model used skip connections to feed low-level features forward to layers near the output. |
| DenseNet | Dense Blocks: 4<br>Dense Layers: 8<br>Growth Rate: 8<br>Dropout Rate: 0.2<br>Parameters: 186,459 | Based on the novel classification model architecture, this model was modified to retain the same image dimension throughout the network. |

*Table 1: Summary of neural network models used in this investigation. The details and important differences are shown here.*

### 9.4.2. *Objective Function*

One of the primary objectives of this investigation was to train a model on data from an existing numerical model, without inheriting all the inaccuracies of that model. The proposed solution presented in this section was to incorporate the OSI-401-b uncertainty field into the objective function calculation so that the model would not be heavily penalized for estimation errors if the concentration label had a high degree of uncertainty. The intention was for the neural network model to learn from features where the existing numerical models are certain. With sufficiently large datasets, the model would be exposed to enough features with a high level of certainty for the overall training process to be successful.

Each model was compiled to use a custom loss function which first separated the label into its two layers so that the concentration and uncertainty patches could be used independently. The loss function then computed the pixel-wise prediction error and scaled it by the pixel-wise "certainty". Figure 10 shows the uncertainty weighted mean absolute error loss function Python definition.

```python
3  def uncertainty_weighted_MAE(y_true, y_pred):
4      # y_true contains the concentration label, and the uncertainty as two two channels
5      y_target    = y_true[:, :, :, :1]
6      uncertainty = y_true[:, :, :, 1:]
7      # Calculate the weighted MAE
8      loss = K.abs(y_pred - y_target)
9      loss = loss * (K.ones_like(loss) - uncertainty) # Scale the error by the 'certainty' of the label
10     return K.mean(loss, axis=-1)
```

*Figure 10: Python definition for the uncertainty weighted mean absolute error function used as the objective function for all models in this investigation. The label passed to the objective function contains both the concentration and uncertainty layers.*

The reasons for defining a custom loss function like this are so that multiple pieces of information can be utilized in determining the loss for each sample, and so that the loss can be constructed as desired for any application. In this case, the estimation error is scaled by the pixel-wise certainty, which prevents the model from learning uncertain details from the existing numerical models as readily as it would without any kind of uncertainty compensation.

### 9.4.3. *Training & Evaluation*

Each model type was compiled using the same custom objective function, and then trained with various combinations of datasets. Table 2 below shows a full list of all models, according to the following naming convention:

*Model_TrainingData_TestData[_DataAugmentation]*

The most interesting results came from using mixed-data training strategies, where the models were partially trained on Northern hemisphere data, and then trained to completion using Southern hemisphere data. These models trained for the same number of epochs as the rest, but exhibited some interesting behavior in their loss trajectories, which is discussed in more detail later on.

Model performance was compared using pure and uncertainty-weighted MAE and MSE. These metrics were tracked throughout the training process, which allows for analysis of loss trajectories as shown in Figure 11, which has the potential to offer more insight than simply looking at the final loss values as shown in Table 2. This discussion is continued in the following section.

For a more subjective analysis, the search interface was used to serve concentration estimations on demand and could be configured to show any combination of model results side-by-side for visual comparison.

## 9.5.    Results & Discussion

### 9.5.1.   *Performance Metrics & Loss Trajectories*

Model performance during and after the training procedure is presented in two ways in this section. Models are compared primarily based on their final performance metrics, but the loss trajectories throughout the training process are also used to gain some insight into each model's behavior. Both the weighted MAE objective function and a pure MAE metric were tracked during all training procedures. Table 2 shows all combinations of model architecture and training strategy used in this investigation and presents the final performance metrics achieved by each model at the end of 50 epochs. It is worth noting that in all cases the weighted MAE scores are less than the pure MAE for each strategy, which confirms that the objective function is indeed penalizing the estimation error less in the presence of uncertainty.

| MODEL NAME | TRAIN DATA | TEST DATA | WEIGHTED MAE (TRAIN, TEST) | MAE (TRAIN, TEST) |
|---|---|---|---|---|
| CNN_N_N | North | North | (0.0640, 0.0697) | (0.0727, 0.0793) |
| CNN_N_N_A | North_A | North | (0.0674, 0.0792) | (0.0763, 0.0897) |
| CNN_S_S | South | South | (0.0579, 0.0660) | (0.0652, 0.0748) |
| CNN_S_S_A | South_A | South | (0.0624, 0.0755) | (0.0703, 0.0855) |
| CNN_NS_S | North + South | South | (0.0561, 0.0631) | (0.0631, 0.0715) |
| CNN_NS_S_A | North_A + South_A | South | (0.0597, 0.0742) | (0.0675, 0.0842) |
| UNet_N_N | North | North | (0.0145, 0.0386) | (0.0166, 0.0437) |
| UNet_N_N_A | North_A | North | (0.0174, 0.0468) | (0.0200, 0.0530) |
| UNet_S_S | South | South | (0.0150, 0.0324) | (0.0170, 0.0369) |
| UNet_S_S_A | South_A | South | (0.0218, 0.0693) | (0.0248, 0.0786) |
| UNet_NS_S | North + South | South | (0.0165, 0.0325) | (0.0187, 0.0370) |
| UNet_NS_S_A | North_A + South_A | South | (0.0198, 0.0477) | (0.0225, 0.0541) |
| DenseNet_N_N | North | North | (0.0603, 0.0797) | (0.0683, 0.0904) |
| DenseNet_N_N_A | North_A | North | (0.0645, 0.124) | (0.0730, 0.138) |
| DenseNet_S_S | South | South | (0.0585, 0.120) | (0.0658, 0.130) |
| DenseNet_S_S_A | South_A | South | (0.0632, 0.130) | (0.0709, 0.141) |
| DenseNet_NS_S | North + South | South | (0.0581, 0.0813) | (0.0654, 0.0897) |
| DenseNet_NS_S_A | North_A + South_A | South | (0.0633, 0.0830) | (0.0711, 0.0919) |

*Table 2: Summary of results obtained from each training strategy for each model type under investigation. Multiple sets of training data indicate that the model was partially trained on the first set, and then trained to completion using the second set.*

Observing the progression of loss metrics throughout the training process can offer some insight into model behavior and may inform decisions when tuning hyperparameters. A particularly interesting set of DenseNet loss trajectories is shown in Figure 11 below, which show a number of interesting features, all relating to the divergence of test results from the decreasing training error curve. The significance of this behavior is discussed in more detail later on.

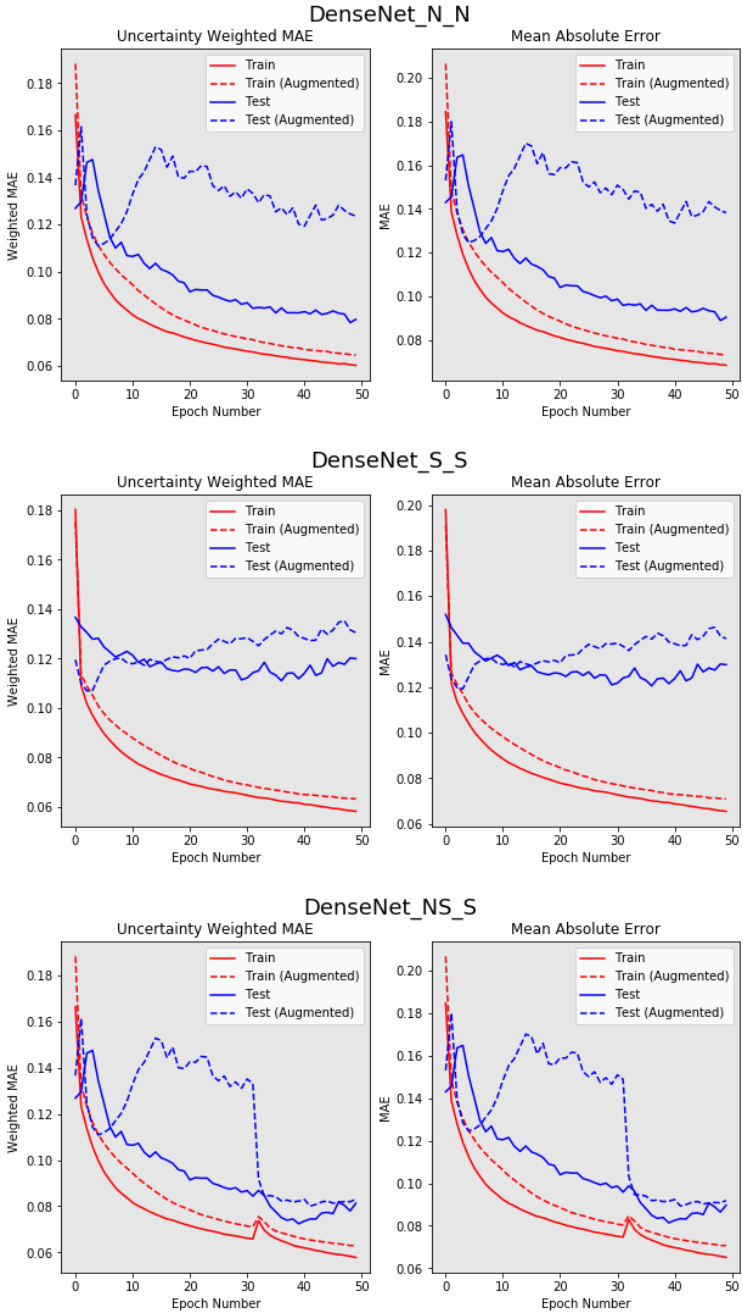

*Figure 11: Loss trajectories for DenseNet, covering all training strategies. The divergence of test results from the training curves is noteworthy and warrants further discussion.*

### 9.5.2. Spatio-Temporal Resolution

One of the primary motivators for choosing to focus on SAR images was the far superior spatial resolution offered by these platforms compared to PMW instruments. This investigation showed that even after the original SAR measurement has been decimated to a much smaller image size, the pixel spacing is still an improvement over PMW data. Since the neural network architectures were designed to replicate the input dimension in their estimates, all of the concentration estimates presented here offer a resolution better than the original data with which they were trained.

The temporal resolution offered by SAR instruments is also an improvement. In many cases, when searching for recent SAR images of a particular location, two images would be returned for the same day, while the most recent PMW-derived estimate was more than a day old. Estimates generated from much more recent SAR data present a clear advantage, particularly in the more dynamic Southern Ocean MIZ where sea ice conditions tend to evolve more quickly.

### 9.5.3. Advantages of Learned Models

The reasons for pursuing a learning approach to generating sea ice concentration estimates were based on the hypothesis that a sufficiently well-designed learning structure would be able to overcome the inaccuracies inherent in the PMW-derived models used for training, even if these inaccuracies are not fully understood. This is similar to how the learned models were able to learn higher-resolution features than were present in any one sample of the training data. This is a non-trivial claim, but the results have shown positive signs indicating that the underlying theory has promise. Some of the more prominent advantages are discussed in detail here.

#### 9.5.3.1. Arbitrary Loss Definition

Perhaps the most useful result of this investigation in the context of trying to create reliable new models using less reliable existing models is the implementation of a certainty-weighted objective function. In the limited scope of this investigation, the weighting relied solely on the uncertainty attribute published with the OSI-401-b concentration product. But this theory can be applied to any arbitrary definition of 'certainty' in future research.

#### 9.5.3.2. Data Augmentation

Data augmentation is a commonly used tool in machine learning. It is particularly useful when dataset size is limited, or to avoid overfitting and in turn help the model generalize better to unseen test data. However, the loss trajectory results presented in the previous section suggest the opposite. In all cases where the use of augmented data was compared with an otherwise identical neural network and training strategy, the test results were worse when the model was trained with augmented data. i.e. the use of augmented data decreased model generalization ability. One possible explanation for this effect is that there may be so little difference between training data and test data that 'overfitting' in training actually yields better test results. However, it should be noted that this is merely conjecture, and any further investigation into this result is out of scope.

### 9.5.3.3.  Multi-Stage Training

It is not uncommon to schedule or dynamically adjust hyperparameters at key points during a model training process, and this is often referred to as multi-stage training. However, in this investigation this staging approach was used to change the entire dataset being used. This has a similar effect to transfer learning, although transfer learning tends to be applied in a cross-domain fashion, whereas in this case both datasets were of sea ice concentration. The rationale for this hybrid training strategy was to allow the model to learn most of its feature associations from Northern hemisphere data, which is known to be more accurate and reliable, before completing the training process with Southern hemisphere data in order to help the model generalize to the new environment. The most interesting observation as a result of this is shown in Figure 11, specifically the DenseNet_NS_S loss trajectory. While training on Northern hemisphere data for the first 32 epochs the train and test scores diverge, which is likely due to overfitting and the fact that the DenseNet model used had far fewer parameters compared to the other two model types (see Table 1). However, almost as soon as the source of training data was switched to the Southern hemisphere after 32 epochs, the test scores rapidly converged with the training scores. As can be seen in Figure 11, this allowed the DenseNet_NS_S strategy to achieve a 4% reduction in weighted MAE on test data compared to the DenseNet_S_S, which both used the same Southern hemisphere test data (see Table 2). This presents another interesting result demonstrating the effects of mixed-data training strategies like this, which could certainly be investigated further in future research.

### 9.5.4.  *Reliability of Existing Models*

As a secondary objective, this investigation aimed to offer some insight into the reliability of the PMW-based OSI-401-b sea ice concentration product used for generating all label information. The comparison tool introduced in Figure 5 allowed for a number of recent in situ observations to be retrospectively compared to the concentration estimates for the same day and location. This comparison revealed significant errors in the concentration estimates within the Southern Ocean MIZ, which illustrates beyond doubt that the naïve application in the Southern hemisphere of estimation models designed and calibrated for the Northern hemisphere will not yield reliable results.

## 9.6.    Conclusions & Future Work

The problem of sea ice concentration estimation is clearly far from solved, although a number of interesting results have been presented over the course of this investigation. Satellite instrument data sources have been under-utilized in the past, but developments in deep learning, particularly in the field of image processing, have led to promising results in this context. The outcome of this investigation is a detailed implementation of an end-to-end workflow for estimating concentration maps from SAR images, based on three neural network models trained on data from existing passive microwave concentration estimation models. The neural network development process was a primary focus of this investigation, and many of the results led to insights on the effects of various data processing techniques used in the course of development, and also pointed to the potential strengths of mixed-data training strategies. Care was taken throughout the development process to validate results and procedures, thus ensuring that the proposed workflow is scalable and robust. The estimation models developed can be improved in many ways but have already shown signs of being able to learn higher resolution features than are contained in any single training sample, and the use of a novel weighted objective function allows the training process to account for known uncertainty within the training data. A comparison tool was presented which allows existing concentration estimates to be compared with in-situ observations and has offered more evidence asserting that the use of Arctic models in the Southern Ocean marginal ice zone will not yield reliable results.

For any future research conducted in this direction, a number of recommendations have already been made, such as defining a more effective 'certainty' metric to be used in the objective function. For example, if one were to generate a 'bias map' using a series of in situ observations for a given region which quantifies any discrepancies between estimates and reality, this bias map could be incorporated into the objective function in a similar way, and may allow the learning process to correct for errors in the training data. One could also investigate the effects of data augmentation in this context which have appeared to behave counterintuitively in these results or propose more effective ways to utilize the benefits of multi-stage and mixed-data training strategies shown here. Additionally, one should consider alternative satellite data sources which may have been overlooked, and also consider scaling up the neural network input dimension to make better use of the high-resolution SAR data which is so freely available.

## 9.7.    References

[1]   A. Alberello, L. Bennetts, P. Heil, C. Eayrs, M. Vichi, K. MacHutchon, M. Onorato and A. Toffoli, "Drift Of Pancake Ice Floes In The Winter Antarctic Marginal Ice Zone During Polar Cyclones," 2019.

[2]   R. Kwok, S. S. Pang and S. Kacimi, "Sea ice drift in the Southern Ocean: Regional patterns, variability, and trends," *Elem Sci Anth,* vol. 5, no. 32, 2017.

[3]   M. Vichi, C. Eayrs, A. Alberello, A. Bekker, L. Bennetts, D. Holland, E. Jong, W. Joubert, K. MacHutchon, G. Messori, J. F. Mojica, M. Onorato, C. Saunders, S. Skatulla and A. Toffoli, "Effects of an Explosive Polar Cyclone Crossing the Antarctic Marginal Ice Zone," *Geophysical Research Letters,* 2019.

[4]   D. Notz, "Challenges in simulating sea ice in Earth System Models," *Wiley Interdisciplinary Reviews: Climate Change,* vol. 3, no. 6, pp. 509-526, 2012.

[5]   S. Häkkinen and D. J. Cavalieri, "Sea ice drift and its relationship to altimetry-derived ocean currents in the Labrador Sea," *Geophysical Research Letters,* vol. 32, no. 11, 2005.

[6]   C. L. V. Cooke and K. A. Scott, "Estimating Sea Ice Concentration From SAR: Training Convolutional Neural Networks With Passive Microwave Data," *IEEE Transactions on Geoscience and Remote Sensing,* vol. 57, no. 7, pp. 4735-4747, 2019.

[7]   G. Huang, Z. Liu and L. van der Maaten, "Densely Connected Convolutional Networks," 2016.

[8]   O. Ronneberger, P. Fischer and T. Brox, "U-Net: Convolutional Networks for Biomedical Image Segmentation," 2015.

[9]   M. de Vos, C.-L. Ramjukadh, C. G. Liesker, M. de Villiers and C. T. Lyttle, "Southern Ocean Marginal Ice Zone Photographs from a 2019 Winter Research Cruise (SA Agulhas II)," 2019. [Online]. Available: https://doi.pangaea.de/10.1594/PANGAEA.905497. [Accessed 5 10 2019].

[10]  K. Żak, "keras-unet," 2019. [Online]. Available: https://github.com/karolzak/keras-unet.

[11]  seasonyc, "densenet," 2018. [Online]. Available: https://github.com/seasonyc/densenet.